\documentclass[twocolumn,showpacs,preprintnumbers,amsmath,amssymb]{revtex4}

\newcommand{\commentoutA}[1]{}

\usepackage{graphicx}
\usepackage{dcolumn}
\usepackage{bm}
\usepackage{algorithm}
\usepackage{algorithmic}

\begin{document}

\preprint{LA-UR 11-02233}

\title{Extended Lagrangian free energy molecular dynamics}

\author{Anders M. N. Niklasson$^{1}$\footnote{Email: amn@lanl.gov}, Peter Steneteg$^{2}$, Nicolas Bock$^{1}$}
\affiliation{$^{1}$Theoretical Division, Los Alamos National Laboratory, Los Alamos, New Mexico 87545}
\affiliation{$^{2}$Department of Physics, Chemistry and Biology (IFM), Link\"{o}ping University, SE-581 83 Link\"{o}ping, Sweden}

\date{\today}

\begin{abstract}
Extended free energy Lagrangians are proposed for first principles 
molecular dynamics simulations at finite electronic temperatures for 
plane-wave pseudopotential and local orbital density matrix based calculations.
Thanks to the extended Lagrangian description the electronic 
degrees of freedom can be integrated by stable geometric schemes that conserve the free energy.
For the local orbital representations both the nuclear and electronic forces
have simple and numerically efficient expressions that are well suited 
for reduced complexity calculations. A rapidly converging recursive Fermi operator 
expansion method that does not require the calculation of eigenvalues and eigenfunctions 
for the construction of the fractionally occupied density matrix is discussed. An efficient
expression for the Pulay force that is valid also for density matrices with fractional
occupation occurring at finite electronic temperatures is also demonstrated. 
\end{abstract}

\maketitle

\section{Introduction}

Molecular dynamics simulations are widely used in materials science, chemistry and molecular biology.
Unfortunately, the most accurate molecular dynamics methods that are based on self-consistent
field (SCF) theory \cite{DMarx00} are often limited by some fundamental shortcomings such as
a very high computational cost or unbalanced phase space trajectories with unphysical hysteresis
effects, numerical instabilities and a systematic long-term energy drift \cite{PPulay04,ANiklasson06}.
Recently an extended Lagrangian framework for first principles molecular dynamics simulations was proposed that
overcomes most of the previous problems \cite{ANiklasson08}. The new framework was originally designed 
for ground state Born-Oppenheimer molecular dynamics
simulations based on Hartree-Fock \cite{Roothaan} or density functional theory \cite{hohen,KohnSham65} 
at {\em zero} electronic temperatures.
In this paper we supplement previous work by proposing extended free energy Lagrangians 
for first principles molecular dynamics simulations also at {\em finite} electronic temperatures \cite{NMermin65,RParr89}.

Our extended Lagrangian free energy dynamics is formulated and demonstrated both for 
plane-wave pseudopotential and local orbital density matrix based calculations.
For the density matrix formulation our approach is given by a generalization
of the original extended Lagrangian framework \cite{ANiklasson08}
using a density matrix representation of the electronic degrees of freedom with fractional occupation of the states.
The plane-wave formulation is based on the wave function extended Lagrangian molecular dynamics method
that recently was proposed by Steneteg et al. \cite{PSteneteg10}.

First we formulate the extended Lagrangian approach to first principles molecular dynamics 
at finite electronic temperatures for local orbital density matrix representations and thereafter
for plane wave pseudopotential calculations. A numerically efficient expression for a generalized Pulay force 
that does not require idempotent density matrices is demonstrated and a recursive
Fermi operator expansion algorithm for the density matrix is discussed. Thereafter a Verlet integration scheme
for the equations of motion is presented.  Finally, we illustrate the extended Lagrangian 
free energy dynamics by some examples before giving a brief summary.

\section{First principles molecular dynamics at finite electronic temperatures}

The first principles molecular dynamics scheme presented in this paper is a finite electronic temperature
generalization of the extended Lagrangian framework of time-reversible Born-Oppenheimer molecular
dynamics \cite{ANiklasson06,ANiklasson08}. In many ways the generalization is straightforward, except
for an additional entropy term, problems arising in the construction of the density matrix, the calculation of
the Pulay force for fractional occupations and the phase alignment problem in the integration of the wave functions
\cite{PSteneteg10}. 
Our two different formulations of extended Lagrangian free energy molecular dynamics are based
on either an underlying local orbital representation or a plane wave pseudo potential description.
Alternative free energy formulations suitable for extended Lagrangian self-consistent tight-binding molecular 
dynamics simulations \cite{ESanville10,GZheng11} will be presented elsewhere.

\subsection{Density matrix extended free energy Lagrangian}

Within an atom-centered local orbital representation we describe our
first principles molecular dynamics at an electronic temperature, $T_e$, 
by a density matrix (DM) extended free energy (XFE) Lagrangian, which we define as 
\begin{equation}\label{XFE}\begin{array}{l}
{\cal L}^{\rm XFE}_{\rm DM} ({\bf R},{\bf \dot R},P,{\dot P}) = \frac{1}{2} \sum_k M_k {\dot R}_k^2 - U[{\bf R};D] \\\
 ~\\
    ~~~~ + \frac{1}{2} \mu \mathrm{Tr}[{\dot P}^2] - \frac{1}{2} \mu \omega^2
    \mathrm{Tr}[(D-P)^2] + T_e{\cal S}[D].
\end{array}
\end{equation}
The extended dynamical variables $P$ and the velocity ${\dot P}$ in
the Lagrangian ${\cal L}^{\rm XFE}_{\rm DM}$ are auxiliary electronic degrees of freedom evolving in a harmonic potential centered around the
self-consistent free energy ground-state solution $D$ \cite{ANiklasson08}. Here $P$, ${\dot P}$,
and the self-consistent free energy ground state $D$, are orthogonal density matrix representations of
the electronic degrees of freedom.  The relation between $D$ and the non-orthogonal (atomic orbital) representation, ${\widetilde D}$,
is given by the congruence transformation
\begin{equation}\label{ZDZ}
{\widetilde D} = ZDZ^T,
\end{equation}
where $Z$ is a congruence factor determined by the overlap matrix $S$
(e.g. see Ref. \cite{ANiklasson07C}) from the condition that
\begin{equation}\label{CTRF}
Z^TSZ = I.
\end{equation}
The constants $\mu$ and $\omega$ are fictitious electron mass and frequency parameters.
The potential $U[{\bf R};D]$ is defined at the self-consistent field (SCF) electron 
ground state $D$, at which the free energy functional 
\begin{equation}\label{Omega}
\Omega[{\bf R},D] = U[{\bf R};D]-T_e{\cal S}[D], 
\end{equation}
has its minimum for a given nuclear configuration, ${\bf R} = \{R_i\}$,
\cite{NMermin65,RParr89} under the constraints of correct electron occupation,
$\mathrm{Tr}[D] = N_{\rm occ}$.  The electronic temperature $T_e$ can be
different from the ionic temperature $T_{\rm ion}$.  We assume that $U[{\bf
R};D]$ is the total electronic energy, including nuclear ion-ion repulsions,
in self-consistent density functional theory, Hartree-Fock theory or some of
their extensions that are based on an underlying SCF description.  The mean
field entropy term in ${\cal L}^{\rm XFE}$, \begin{equation}\label{SED} {\cal
S}[D] = -2 k_B \mathrm{Tr}[D\ln(D) + (I-D)\ln(I-D)], \end{equation} is
included to provide the variationally correct energetics and dynamics
\cite{MWeinert92,RWentzcovitch92}. The factor 2 above is used for restricted
calculations (not spin polarized) when each state is doubly occupied, which is assumed throughout
the paper.

\subsubsection{Euler-Lagrange equations of motion}

The time evolution of the system described by ${\cal L}^{\rm XFE}_{\rm DM}$ is determined by the Euler-Lagrange equations,
\begin{eqnarray}
\frac{d}{dt} \left(\frac{\partial {\cal L}^{\rm XFE}_{\rm DM}}{\partial {\dot
R}_k}\right) & = & \frac{\partial  {\cal L}^{\rm XFE}_{\rm DM}}{\partial {R}_k},\\
\frac{d}{dt} \left(\frac{\partial {\cal L}^{\rm XFE}_{\rm DM}}{\partial {\dot P}}\right) & = & \frac{\partial  {\cal L}^{\rm XFE}_{\rm DM}}{\partial P},
\end{eqnarray}
which give the equations of motion for the nuclear and electronic degrees of freedom,
\begin{eqnarray}
\label{EL_a}
M_k {\ddot R}_k & = & -\frac{\partial U\left[{\bf R};D\right]}{\partial R_k} + 
   T\frac{\partial {\cal S}[D]}{\partial R_k} - \frac{\mu\omega^2}{2}
   \frac{\partial \mathrm{Tr}[(D-P)^2]}{\partial R_k}, \nonumber \\
\\
\label{EL_b}
\mu {\ddot P} & = & \mu {\omega^2} (D - P).
\end{eqnarray}
In the limit $\mu \rightarrow 0$, we get the decoupled equations
\begin{eqnarray}
\label{EL_1}
M_k {\ddot R}_k & = & -\frac{\partial U\left[{\bf R};D\right]}{\partial R_k} +
T_e\frac{\partial {\cal S}[D]}{\partial R_k}, \\
\label{EL_2}
{\ddot P} & = & {\omega^2} (D - P),
\end{eqnarray}
where the self-consistent density matrix $D$ is given from the Fermi operator expansion
\begin{equation}\label{FO}
D = \left[e^{\beta(H-\mu_0 I)} + I\right]^{-1},
\end{equation}
under the canonical condition of charge neutrality, 
\begin{equation}
\mathrm{Tr}[D] = \mathrm{Tr}[{\widetilde D}S] = N_{\rm occ.},
\end{equation}
i.e. the chemical potential $\mu_0$ is set such that the trace of $D$ is the number of occupied states. 
The inverse temperature $\beta = 1/(k_B T_e)$. The orthogonalized effective single-particle Hamiltonian $H$, i.e. the Fockian or 
Kohn-Sham Hamiltonian, is given by a congruence transformation of the Fockian in a non-orthogonal (atomic orbital)
representation, ${\widetilde H}$, where
\begin{equation}
H = Z^T{\widetilde H}Z.
\end{equation}
Since we use the limit $\mu \rightarrow 0$, we recover the regular free energy Lagrangain (without extention) and 
the total free energy is a constant of motion that is unaffected by the extended variables $P$ and ${\dot P}$.
This is in contrast to extended Lagrangian Car-Parrinello type molecular dynamics \cite{RCar85,BHartke92,HBSchlegel01,JMHerbert04}.
The major advantage of the auxiliary set of variables $P$ and ${\dot P}$ occurs in the SCF
optimization and in the geometric integration of the equations of motion discussed in section III below.

\subsection{Wave Function Extended Free Energy Lagrangian}

In plane wave (PW) pseudopotential methods it is convenient to propagate both the electronic
wave functions, $\Phi = \{\phi_{m}({\bf r})\}$, and the density, $\rho({\bf r})$. In this case we define the wave function (and density) extended
free energy Lagrangian \cite{PSteneteg10} as
\begin{equation}\label{XFE_PW}\begin{array}{l}
{\cal L}^{\rm XFE}_{\rm PW} ({\bf R},{\bf \dot R},\Phi, {\dot \Phi},\rho , {\dot \rho}) 
= \frac{1}{2} \sum_k M_k {\dot R}_k^2 - U[{\bf R};n^{\rm sc}] \\\
 ~\\
   + \frac{1}{2} \mu \sum_{m}\int |{\dot \phi}_{m}({\bf r})|^2 d{\bf r} 
             - \frac{1}{2} \mu \omega^2 \sum_{m}\int |\psi^{sc}_{m}({\bf r}) - \phi_{m}({\bf r})|^2 d{\bf r}  \\
 ~\\
   + \frac{1}{2} \mu \int |{\dot \rho({\bf r})}|^2 d{\bf r} 
             - \frac{1}{2} \mu \omega^2 \int |n^{sc}({\bf r}) - \rho({\bf r})|^2 d{\bf r} 
   + T_e{\cal S}[{\bf f}].
\end{array}
\end{equation}
The auxiliary electronic degrees of freedom $\Phi$ and $\rho$ are
extended through harmonic oscillators centered around the self-consistent (sc) ground state wave functions
$\Psi^{\rm sc} = \{\Psi_m({\bf r})\}$ and densities $n^{\rm sc}$.
As above, $\mu$ is a fictitious electron mass parameter and $\omega$ is a frequency parameter determining the curvature of the harmonic
potentials. 
As for the density matrix extension, the potential $U[{\bf R};n^{\rm sc}]$ is defined at the self-consistent field (SCF) electron
ground state $n^{\rm sc}$, at which the free energy functional
\begin{equation}\label{Omega_PW}
\Omega[{\bf R},n^{\rm sc}] = U[{\bf R};n^{\rm sc}]-T_e{\cal S}[{\bf f}],
\end{equation}
has its minimum for a given nuclear configuration, ${\bf R} = \{R_i\}$, \cite{NMermin65,RParr89} under
the constraints of correct electron occupation, $\int n^{sc}({\bf r}) d{\bf r} = N_{\rm occ}$.
The mean field entropy term in ${\cal L}^{\rm XFE}_{\rm PW}$ is given analogous to Eq.\ (\ref{SED})
as a function of the occupation numbers, ${\bf f} = \{f_i\}$, where
\begin{equation}\label{S_PW}
{\cal S}[{\bf f}] = -2 k_B \sum_i\left(f_i\ln(f_i) + (1-f_i)\ln(1-f_i)\right),
\end{equation}
and
\begin{equation}\label{Occ_Nr}
f_i = \left[e^{\beta(\varepsilon_i-\mu_0)} + 1\right]^{-1}.
\end{equation}
Here $\varepsilon_i$ are the eigenvalues of the effective single particle (Kohn-Sham) Hamiltonian, i.e.
\begin{equation}
H\psi_i = \varepsilon_i \psi_i.
\end{equation}
The entropy term ${\cal S}[{\bf f}]$ provides a variationally correct energetics and dynamics
\cite{MWeinert92,RWentzcovitch92} where the electron density is given by
\begin{equation}\label{density}
n^{\rm sc}({\bf r}) = \sum_m \int f_m |\psi^{\rm sc}_m({\bf r})|^2 d{\bf r}. 
\end{equation}

\subsubsection{Euler-Lagrange equations of motion}

As for the density matrix extended free energy Lagrangian, the dynamics for the wave function extended free energy Lagrangian is
determined by the Euler-Lagrange equations, which give the equations of motion
for the nuclear and electronic degrees of freedom,
\begin{eqnarray}
\label{EL_a_PW}
& & M_k {\ddot R}_k = -\frac{\partial U\left[{\bf R};n^{\rm
sc}\right]}{\partial R_k} + \\
& & ~~ {} - \frac{\mu\omega^2}{2} \frac{\partial}{\partial R_k}
\left(\sum_{n}\int |\psi^{sc}_{n} - \phi_{n}|^2 d{\bf r} \right) \nonumber \\
& & ~~ {} - \frac{\mu\omega^2}{2} \frac{\partial}{\partial R_k} \left(\int
|n^{\rm sc}({\bf r}) - \rho({\bf r})|^2 d{\bf r}\right) + T_e\frac{\partial
{\cal S}[{\bf f}]}{\partial R_k}, \nonumber \\
\label{EL_b_PW}
& & \mu {\ddot \Phi} = \mu {\omega^2} (\Psi^{\rm sc} - \Phi), \\
\label{EL_c_PW}
& & \mu {\ddot \rho}({\bf r}) = \mu {\omega^2} (n^{\rm sc}({\bf r}) -
\rho({\bf r})).
\end{eqnarray}
In the limit $\mu \rightarrow 0$, we get the decoupled equations
\begin{eqnarray}
\label{EL_1_PW}
M_k {\ddot R}_k & = & -\frac{\partial U\left[{\bf R};n^{\rm
sc}\right]}{\partial R_k} + T_e\frac{\partial {\cal S}[{\bf f}]}{\partial
R_k}, \\
\label{EL_2_PW}
{\ddot \Psi} & = & {\omega^2} (\Psi^{\rm sc} - \Phi), \\
\label{EL_3_PW}
{\ddot \rho}({\bf r}) & = & {\omega^2} (n^{\rm sc}({\bf r}) - \rho({\bf r})).
\end{eqnarray}
Also in this case, the total free energy is a constant of motion in the limit $\mu \rightarrow 0$.
The advantages with the extended equations of motion occur in the SCF optimization
and geometric integration discussed in section III.

\subsection{Generalized Pulay force}

The nuclear force ${\cal F}^{\rm tot} = M_k {\ddot R}_k$ in Eqs.\ (\ref{EL_1}) and (\ref{EL_1_PW}) can be partitioned into three separate terms,
\begin{equation}\label{HFF}
{\cal F}^{\rm tot} = {\cal F}^{\rm HF} + {\cal F}^{\rm P} + {\cal F^S}.
\end{equation}
The first term, ${\cal F}^{\rm HF}$, is the Hellmann-Feynman force
term \cite{PPulay69,HSchlegel00} for a fractional occupation of
the states \cite{MWeinert92,RWentzcovitch92} (including nuclear-nuclear repulsion). 
The second term, ${\cal F}^{\rm P}$,
corresponds to the Pulay force term \cite{PPulay69}, and the third term,
\begin{equation}
{\cal F^S}_k = T_e({\partial}{\cal S}/{\partial R_k}),
\end{equation}
is an additional entropy force term.

In a plane wave formulation the Pulay force term usually vanishes, but only at $T_e = 0$.
At finite electronic temperatures the Pulay term is finite even for a constant plane wave basis set representation.
In this case however, the Pulay term is exactly cancelled by the entropy force term ${\cal F^S}$ 
if the entropy is given by Eq.\ (\ref{S_PW}). This is not true for a local orbital basis set representation.
In this case only a part of the Pulay force term ${\cal F}^{\rm P}$ is canceled by the entropy contribution ${\cal F^S}$
\cite{ANiklasson08b}. Nevertheless, this cancellation provides a significant simplification of the Pulay term. The remaining force term can 
be viewed as a generalized ($G$) Pulay force \cite{ANiklasson08b},
\begin{equation}\label{GPForce}
{\cal F}^{\rm P}_G = {\cal F}^{\rm P} + {\cal F^S} = 2 \mathrm{Tr}[S^{-1}{\widetilde H}{\widetilde D}(\partial S/\partial R_k)],
\end{equation}
which is valid for finite temperature ensembles with non-integer occupation \cite{ANiklasson08b}.
For a constant orthonormal plane wave basis set representation the overlap matrix is equal to the identity matrix, i.e. $S = I$,
and the generalized Pulay term disappears.
A similar convenient expression for the Pulay force term was recently given by Schlegel et al. \cite{HBSchlegel01}, 
but without considering any entropy contribution from fractional occupation. A generalized Pulay force was 
also discussed in connection to the efficient Ehrenfest-like molecular dynamics scheme by Jakowski and Morokuma \cite{JJakowski09}. 
In the present formulation the electronic entropy is necessary to make our free energy and forces variationally 
correct in a molecular dynamics simulation with
factional occupation of the states at finite electronic temperatures \cite{MWeinert92,RWentzcovitch92}. The density matrix or the density that
minimizes the free energy functional in Eq.\ (\ref{Omega}) or in Eq.\ (\ref{Omega_PW}) 
with the particular form of the entropy term in Eq.\ (\ref{SED}) or in Eq.\ (\ref{S_PW})
are given by the Fermi operator expansion in Eq.\ (\ref{FO}) or by Fermi
weighted integration in Eq.\ (\ref{density}). 
If other expressions of the entropy are used, 
the Pulay force expression in Eq.\ (\ref{GPForce}) may still be valid \cite{RWarren96}, but the density matrix $D$ 
and wave functions will 
have a fractional occupation distribution of the eigenvalues that is different from the Fermi-Dirac function.

\section{Geometric integration}\label{Int}

Thanks to the extended Lagrangian formulations of free energy molecular dynamics, Eqs.\ (\ref{XFE}) and (\ref{XFE_PW}), it is possible
to simultaneously integrate both the nuclear and the electronic degrees of freedom in Eqs.\ (\ref{EL_1}) and (\ref{EL_2})
for the density matrix formulation,
or Eqs.\ (\ref{EL_1_PW}) and (\ref{EL_2_PW}) for the plane wave extension,
with almost any kind of geometric integration scheme developed for celestial or classical molecular dynamics, 
e.g. the time-reversible Verlet algorithm
or higher-order symplectic integration methods \cite{BJLeimkuhler04,ANiklasson08,AOdell09,PSteneteg10}.
Morevover, the auxiliary degrees of freedom $P$ in Eq.\ (\ref{XFE}) or $\Phi$ and $\rho$
in Eq.\ (\ref{XFE_PW}) will stay close to the free energy ground state $D$ or
$\Psi^{\rm sc}$ and $n^{\rm sc}$, since $P$, $\Phi$, and $\rho$ evolve in
harmonic potentials centered around $D$, $\Psi^{\rm sc}$, and $n^{\rm sc}$.  It
is therefore possible to use the extended dynamical variables as an efficient
initial guess of $D$, $\Psi^{\rm sc}$, or $n^{\rm sc}$, in the iterative SCF
optimization procedure that minimizes the free energy, i.e. 
\begin{eqnarray}
D(t) = {\rm SCF}\left[{\bf R};P(t)\right],\\
~~ \nonumber \\
\{\Psi^{\rm sc} (t),n^{\rm sc} (t)\} = {\rm SCF}\left[{\bf R};\Phi(t),\rho(t)\right].
\end{eqnarray}
This choice of initial guesses also avoids the fundamental problem of a broken time-reversal symmetry
in the propagation of the underlying electronic degrees of freedom \cite{PPulay04,ANiklasson06,ANiklasson08}, 
since the extended dynamical variables
can be integrated by a time-reversible algorithm, for example, the Verlet scheme \cite{LVerlet67}. 
This is in contrast to regular Born-Oppenheimer-like molecular dynamics, where the initial guess
of the SCF optimization is based on some extrapolation of the self-consistent solutions from previous time steps 
\cite{MCPayne92,TAArias92,JMMillan99,PPulay04,CRaynaud04,JMHerbert05}, for example 
\begin{eqnarray}
\label{EXTRP}
D(t)  =  {\rm SCF}\left[{\bf R}; 2D(t-\delta t) - D(t-2\delta t)\right],  \\
~~ \nonumber \\
\{\Psi^{\rm sc} (t),n^{\rm sc} (t)\} = {\rm SCF}\left[{\bf R};\Phi(t-\delta t),\rho(t-\delta t)\right],
\end{eqnarray}
which breaks time-reversal symmetry because of the incomplete and non-linear SCF optimization.
A broken time-reversal symmetry leads to a hysteresis effect and a systematic drift in the total energy and the phase space
\cite{PPulay04,ANiklasson06}.  Only by the costly procedure of increasing the SCF convergence to a very high degree 
of accuracy is it possible to reduce the drift, though it never really disappears. As will be shown below, our
extended Lagrangian approach allows for highly efficient and stable free energy molecular dynamics simulations without any significant
energy drift. Often only one single SCF iteration is sufficient to provide accurate trajectories.

In principle, time-reversibility is not a necessary criterion for long-term energy conservation. A more
fundamental aspect is the conservation of geometric properties of the exact flow of the underlying dynamics 
as in symplectic integration schemes \cite{BLeimkuhler04,ANiklasson08,AOdell09}. 
Here we only consider the Verlet integration algorithm, which in its velocity formulation is both
time-reversible and symplectic. First we present the integration of the density matrix and thereafter
the electron wave function and density integration.

\subsection{Density matrix integration} 

The time-reversible Verlet integration \cite{LVerlet67} of the density matrix $P(t)$ in Eq.\ (\ref{EL_2}),
\begin{equation}\label{VRL}
P(t+\delta t) = 2P(t) - P(t-\delta t) + \delta t^2 {\ddot P}(t),
\end{equation}
that includes a weak external dissipative force that removes the accumulation of numerical noise \cite{ANiklasson09},
has the following form:
\begin{equation}\label{VRL_Damp}
P_{n+1} = 2P_{n} - P_{n-1} + \delta t^2 \omega^2 (D_n - P_n) + \alpha \sum_{k=0}^K c_k P_{n-k}.
\end{equation}
In the first $K+1$ initial steps we set $P_n = D_n$, where $D_n = D(t_0 + n\delta t)$, and we use
a high degree of SCF convergence, typically 10-15 SCF cycles per time step.
The last term on the right hand side of Eq.~(\ref{VRL_Damp}) is the dissipative force term, which is tuned by some small
coupling constant $\alpha$. This term corresponds to a weak coupling to an external system that 
removes numerical error fluctuations, but without any significant modification of the microcanonical
trajectories and the free energy \cite{ANiklasson09}. 
A few optimized examples of the dimensionless parameter, $\delta t^2 \omega^2$,
the $c_k$ coefficients and $\alpha$ are given in Table~\ref{Tab_Coef} (see Ref.\ \cite{ANiklasson09}
for details).  The integration coefficients in Table~\ref{Tab_Coef} are optimized under the condition of
stability under incomplete, approximate SCF convergence.
A larger value of $\alpha$ in Table~\ref{Tab_Coef} gives more dissipation but also
a larger perturbation of the molecular trajectories. For $\alpha = 0$
the scheme is exactly time-reversible and perfectly lossless without any dissipation of numerical errors.
In an exactly time-reversible and lossless scheme small numerical 
errors may accumulate to large fluctuations during long simulations, which can
lead to a loss of numerical stability. 
It is particularly important to include dissipation when the numerical noise is large, 
for example, in reduced complexity calculations utilizing an approximate sparse matrix algebra. 

\begin{table}[t]
  \centering
  \caption{\protect Coefficients for the Verlet integration scheme with the external dissipative
  force term in Eq.\ (\ref{VRL_Damp}). The values are derived in Ref.\ \cite{ANiklasson09}, which
  contains a more complete set of coefficients.
  }\label{Tab_Coef}
  \begin{ruledtabular}
  \begin{tabular}{llccccccccc}
    $K$ & $\delta t^2 \omega^2$ & \!\!$\alpha\!\times\!10^{-3}$\!\! & \!$c_{0}$ & $c_{1}$ & $c_{2}$ & $c_{3}$ & $c_{4}$ &
$c_{5}$ & $c_{6}$ & $c_{7}$ \\
     \hline
     0  & 2.00 & \!\!0\!\! &   &    &   &  &  &      &     &     \\
     3  & 1.69 & \!\!150\!\! & -2    &   3   &   0  & -1  &     &      &     &     \\
     5  & 1.82 & \!\!18\!\! &-6    &   14   &  -8  & -3  &  4  &  -1  &     &      \\
     7  & 1.86 & \!\!1.6\!\! &-36    &   99  & -88  & 11  &  32 & -25  & 8   &  -1  \\
  \end{tabular}
  \end{ruledtabular}
\end{table}

\subsection{Wave function and density integration}

The Verlet integration of the wave functions and the density in Eqs.\ (\ref{EL_2_PW}) and (\ref{EL_3_PW}), including a weak external
dissipative force term that removes a systematic accumulation of numerical noise (as described above)
has the following form:
\begin{eqnarray}
\label{Verlet_PW}
\Phi_{n+1} & = & 2\Phi_n - \Phi_{n-1} + \delta t^2 \omega^2 (\Psi^{\rm sc}_n
U_n - \Phi_n) \nonumber \\
& & {} + \alpha \sum_{k = 0}^{K} c_k \Phi_{n-m}, \\
\rho_{n+1} & = & 2\rho_n - \rho_{n-1} + \delta t^2 \omega^2 (n^{\rm sc}_n -
\rho_n) \nonumber \\
& & {} + \alpha \sum_{k = 0}^{K} c_k \rho_{n-m}.
\end{eqnarray}
Sine the electronic ground state wave functions, $\Psi_n^{\rm sc}$, are unique except with respect to their phase, a straightforward
Verlet integration of the wave functions may lead to large and uncontrolled errors. To avoid this problem we have included
a subspace alignment through a unitary rotation transform $U_n$ in Eq.\
(\ref{Verlet_PW}) \cite{PSteneteg10}, where
\begin{equation}
U_n = \arg~ \min_{U'} \|\Psi_n U' - \Phi_n\|_F,
\end{equation}
where $\|\cdot\|_F$ denotes the Frobenius norm. The rotation $U_n$ that minimizes this distance between
$\Psi_n U'$ and $\Phi_n$ is given by
\begin{equation}
U = (OO^{\dagger})^{-1/2}O,
\end{equation}
where $O = \langle \Psi^{\rm sc}_n|\Phi_n \rangle$ \cite{RevModPhys.64.51}.
Using this subspace alignment we can use the same coefficients as in the
density matrix integration that are given in Tab.\ \ref{Tab_Coef}.

\section{Recursive Fermi operator expansion for the density matrix}

The density matrix can be calculated from a diagonalization of the effective single-particle Hamiltonian, $H$, where
\begin{equation}
Q E_{\varepsilon} Q^T = H.
\end{equation}
Here $Q$ is an orthogonal matrix consisting of the Hamiltonian eigenvectors and $E_{\varepsilon}$ is a diagonal matrix with
the corresponding Hamiltonian eigenvalues $\{ \varepsilon_i \}$. By replacing these eigenvalues $\{ \varepsilon_i \}$
with the occupation numbers $\{f_i\}$ in Eq.\ (\ref{Occ_Nr}), the density matrix is given by
\begin{equation}
P = Q E_{f} Q^T.
\end{equation}
Because of the cubic scaling of the computational cost as a function of system size, 
the diagonalization approach is ill-suited for reduced complexity calculations of large extended systems.
In this case we instead propose using recursive Fermi operator expansion methods, both
at finite and zero electronic temperatures. In plane wave pseudopotential methods usually some
form of iterative eigensolvers are used to calculate a subset of the lowest lying eigenfunctions \cite{MCPayne92}.
Plane wave matrix representations are not sparse and linear scaling of the cost as a function
of the number of atoms can in general not be achieved. Because of the large number of plane wave basis
functions, an explicit construction of the density matrix is difficult and it is only for
local atom centered basis set representations for which it is useful to calculate the density matrix.
Here we will discuss efficient recursive expansion methods for the construction of the
density matrix at finite and zero electronic temperatures. The recursive Fermi operator expansion
methods allows for very efficient reduced complexity calculations of large systems if sparse matrix
algebra can be used.

\subsection{$T_e \ge 0$}

The density matrix at finite electronic temperatures used in the calculation of the nuclear 
and electronic forces, Eqs.\ (\ref{EL_1}) and (\ref{EL_2}), can be constructed by a recursive Fermi operator expansion \cite{ANiklasson03B},
\begin{equation}\label{Rec}
D = \left[e^{\beta(H-\mu_0 I)} + I\right]^{-1} \approx
{\cal P}_m({\cal P}_{m-1}(\ldots {\cal P}_0^{(m)}(H)\ldots )).
\end{equation}
There are certainly numerous fairly well established alternative techniques that can incorporate
the effects of the electronic distribution at finite temperatures, such as Matsubara,
Green's function, integral representation, Chebyshev Fermi operator expansion, or continued fraction methods
\cite{TMatsubara55,Mahan81,SGoedecker93,NBernstein01,SGoedecker94,SGoedecker95B,RSilver96,SGoedecker99,WLiang03,TOzaki07,MCeriotti08}.
However, in contrast to schemes that are based on {\em serial} expansions, 
the {\em recursive} expansion in Eq.\ (\ref{Rec}) leads
to a very high-order approximation in only a few number of iteration steps. 
Here we will use a recursive Fermi operator expansion based on the Pad\'{e} polynomial 
\begin{equation}
{\cal P}_n(x)  = x^2/[x^2 + (1-x)^2]
\end{equation}
in Eq.\ (\ref{Rec}) for $n>0$ \cite{ANiklasson03B}.

The problem of finding the chemical potential $\mu_0$ in Eq.\ (\ref{Rec}) such that the correct occupation is achieved,
\begin{equation}
\mathrm{Tr}(D) = N_{\rm occ},
\end{equation}
can be solved by a Newton-Raphson optimization using the analytic derivative
of the density matrix with respect to $\mu_0$,
\begin{equation}
  \frac{\partial D}{\partial \mu_0} = \beta D (I-D).
\end{equation}

The recursive Fermi operator expansion of the temperature-dependent density
matrix, which automatically finds the chemical potential by a Newton-Raphson
optimization, is given by Algorithm~\ref{Fermi_Op_Exp}.  The algorithm is a
slight modification of the original scheme given in Ref.~\cite{ANiklasson03B}
and was previously presented in Ref.\ \cite{ANiklasson08b}. A full derivation
is given for the first time in
Appendix~\ref{appendix:recursive_operator_expansion}.  The main difference to
the original algorithm is that the expansion order $m$ is independent of the
temperature and that the expansion is performed for $D$ directly, instead of
the virtual projector $I-D$.  Note also that the choice of $m$ determines both
the initialization of $X_0 = {\cal P}_0^{(m)}(H)$ as well as the expansion order as is shown in the
appendix. The initial guess for $\mu_0$ in the Newton-Raphson optimization has
to be particularly good at low temperatures, i.e. when $\beta$ is large. If 
$m$ is small the analytic derivative $\frac{\partial D}{\partial \mu_0} =
\beta D (I-D)$ is more approximate, which also may require a better initial guess.

\begin{algorithm}
\caption{Recursive Fermi operator expansion, \cite{ANiklasson03B,ANiklasson08b}}
\label{Fermi_Op_Exp}
\algsetup{indent=2em}
\begin{algorithmic}
\STATE $H \gets Z^T{\widetilde H}Z$
\STATE $m \gets \mbox{Number of recursive iterations}$
\STATE $\beta \gets 1/(k_B T_e)$
\STATE $\mu_0 \gets \mbox{Initial guess of } \mu_0$
\WHILE{Occupation Error $>$ Tolerance}
  \STATE $X_0 \gets \frac{1}{2}I - (H-\mu_0 I)\beta/2^{2+m}$
  \FOR{$n = 1:m$}
    \STATE solve $[X_{n-1}^2 + (I-X_{n-1})^2]X_n = X_{n-1}^2$
  \ENDFOR
  \STATE $D \gets X_m$
  \STATE Occupation Error $= | \mathrm{Tr}(D) - N_{\rm occ}|$
  \STATE $\mu_0 \gets \mu_0 + [N_{\rm occ} - \mathrm{Tr}(D)]/\mathrm{Tr}[\beta D (I-D)]$
\ENDWHILE
\STATE ${\widetilde D} \gets ZDZ^T$
\end{algorithmic}
\end{algorithm}

In the recursive Fermi operator expansion, Algorithm~\ref{Fermi_Op_Exp}, a
system of linear equations is solved in each iteration.  The numerical problem
is well conditioned and the solution from the previous cycle $X_{n-1}$ is
typically close to the solution $X_n$.  A linear conjugate-gradient solver is
therefore very efficient \cite{ANiklasson03B,GGolub96}.  The Fermi operator
expansion algorithm can be formulated based only on matrix-matrix operations
and does not require a diagonalization of the Fockian. It is therefore
possible to reach a reduced complexity scaling of the computational cost as a
function of system size if sparse matrix algebra can be utilized.

\subsection{$T_e = 0$}

The recursive Fermi operator expansion, Algorithm~\ref{Fermi_Op_Exp}, provides
an efficient calculation of the density matrix at finite electronic
temperatures, $T_e \ge 0$. The algorithm can be applied also at zero
temperatures, $T_e = 0$. However, in this case there are more efficient
recursive Fermi operator expansion schemes, for example, the second order 
occupation correcting spectral projection technique \cite{ANiklasson02}, shown in
Algorithm~\ref{Fermi_Op_Exp_T0}.  Here $\sigma_n = \pm 1$ and is chosen to
minimize the occupation error, $|\mathrm{Tr}[X_{n+1}] - N_{\rm occ}|$.  The
idempotency error,
\begin{equation}\label{SP2}
{\rm Idempotency~Error} = |\mathrm{Tr}[X_n-X_n^2]| = |\mathrm{Tr}[X_n(I-X_n)]|,
\end{equation}
which determines the degree of convergence, 
can be estimated from the size of the occupation correction, 
since $|\mathrm{Tr}[X_n(I-X_n)]| = |\mathrm{Tr}[X_{n+1}] - \mathrm{Tr}[X_n]|$,
and towards convergence (e.g. $n > 10$) the continuous decrease of the idempotency 
error can be estimated from every second value.  In the
initialization, $\varepsilon_{\rm max}$ and $\varepsilon_{\rm min}$ are upper
and lower estimates of the spectral bounds of $H$. To improve matrix sparsity,
numerical thresholding can be applied after each matrix multiplication within
controllable total error bounds
\cite{ANiklasson03,EHRubensson05,ANiklasson07C,EHRubensson08}. The recursive 2nd-order occupation
correcting expansion algorithm is fast, robust, and very memory efficient
compared to other linear scaling methods \cite{MDaw93,LNV,APalser98,EHRubensson11}.
If more detailed information about the homo-lumo gap and the spectral bounds
are available, an efficient boosting technique based on shifting and rescaling
of the second order projection polynomials in Eq.\ (\ref{SP2}) can be applied 
to accelerate the idempotency convergence \cite{EHRubensson11b}.

\begin{algorithm}
\caption{Recursive Fermi operator expansion at zero electronic temperature \cite{ANiklasson02}}
\label{Fermi_Op_Exp_T0}
\algsetup{indent=2em}
\begin{algorithmic}
\STATE $H \gets Z^T{\widetilde H}Z$
\STATE $X_0 \gets (\varepsilon_{\rm max} - \varepsilon_{\rm min})^{-1}(\varepsilon_{\rm max}I-H)$
\WHILE{the Idempotency Error is decreasing}
  \STATE $X_{n+1} \gets X_n + \sigma_n(X_n-X_n^2)$
\ENDWHILE
\STATE ${\widetilde D} \gets \lim_{n \rightarrow \infty} ZX_nZ^T$
\end{algorithmic}
\end{algorithm}

\section{Examples}

\subsection{Computational details}

The extended Lagrangian free energy molecular dynamics schemes were implemented in FreeON,
as suite of freely available programs for {\em ab initio} electronic structure calculations developed 
by Challacombe and co-workers \cite{FreeON}, and in VASP, the Vienna Ab initio Simulation Package \cite{GKresse96}. 

For the calculation of the electronic entropy
contribution to the free energy we rely on a diagonalization of the Fockian or 
the Kohn-Sham Hamiltonian, which would not be possible within a reduced complexity scheme. 
A sufficiently accurate expansion of $\cal{S}$ in Eq.\ (\ref{SED}), which is suitable within a 
reduced complexity calculation may be possible, but it is currently 
an unsolved problem.  For the construction of the density matrix we use either the recursive Fermi operator 
expansion, Algorithm~\ref{Fermi_Op_Exp}, or, as a comparison, an ``exact'' calculation based on the conventional (non-recursive) 
exponential form of the Fermi-Dirac distribution in Eq.\ (\ref{Rec}), which requires a full diagonalization of the Hamiltonian.

The plane wave pseudo potential calculations were carried out 
within the local density approximation (LDA) for the exchange correlation energy. Ultrasoft pseudo potentials were used \cite{Vanderbilt:1990p7892},
and for the SCF optimization an iterative RMM-DIIS scheme \cite{PPulay80,Wood:1985p5116} was applied.  The plane wave cutoff was 246 eV and a mesh of
64 {\it k}-points was used. The electronic temperature was set to $T_e = 500$ K around which also the nuclear temperature fluctuated.

\subsection{Free energy conservation}

\subsubsection{Density matrix extension}

\begin{figure}[t]
\includegraphics[width=\columnwidth]{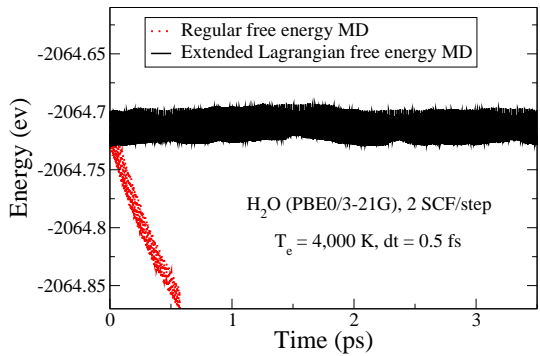}
\caption{\label{E_Drift}
\small (Color online) The fluctuations in the total free energy, $E^{\rm tot}_F = E_{\rm K} + U - T_e{\cal S}$, 
of a water molecule using regular free energy molecular dynamics for a conventional density matrix based
Born-Oppenheimer method with a linear extrapolation of the electronic state, Eq.\ (\ref{EXTRP}), 
in comparison to the density matrix extended Lagrangian free energy molecular dynamics.}
\end{figure}

In a free energy simulation based on regular first principles molecular
dynamics the self-consistent
electronic state is propagated through an extrapolation from previous time
steps, which is used as an initial guess to the SCF optimization procedure as
was shown in Eq.\ (\ref{EXTRP}).  This approach leads to the breaking of
time-reversal symmetry with an unphysical systematic drift in the energy and
phase space. Figure \ref{E_Drift} illustrates this problem, which is avoided
with the density matrix extended Lagrangian free energy molecular dynamics.

\begin{figure}[t]
\includegraphics[width=\columnwidth]{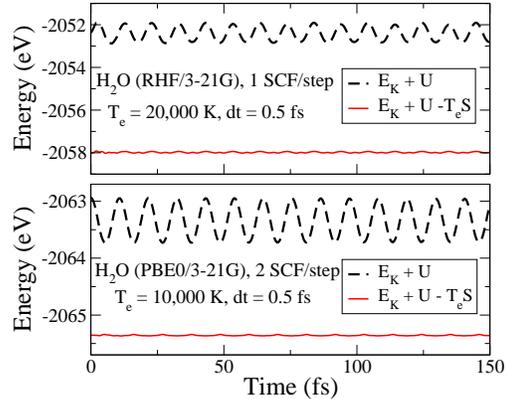}
\caption{\label{Fig_water_T10000K}
\small (Color online) The fluctuations in the nuclear kinetic and potential energy, $E_{\rm K} + U$, in comparison
to the total free energy, $E^{\rm tot}_F = E_{\rm K} + U - T_e{\cal S}$, for a density matrix extended Lagrangian free energy
molecular dynamics simulation of a single water molecule using Hartree-Fock (RHF) or
hybrid density functional theory (PBE0 \cite{CAdamo99}). The average nuclear temperature was approximately
room temperature, i.e. $T_{\rm ion}\approx 300$ K.}
\end{figure}

Figure \ref{Fig_water_T10000K} shows the free energy conservation vs.
the fluctuations of the sum of the nuclear kinetic and potential energy,
$E_{\rm K} + U$. Only one or two SCF cycles per time step are used, which
shows that the simulation is stable, conserving the free energy under
incomplete approximate SCF convergence. While the amplitude of the fluctuations in $E_{\rm
K} + U$ is dependent on the electronic temperature $T_e$, the amplitude of the
total free energy oscillations, $E^{\rm tot}_F = E_{\rm K} + U - T_e{\cal S}$,
is not. This is an agreement with the total free energy being a constant of motion as
in the density matrix extended free energy Lagrangian, Eq.\ (\ref{XFE}), in
the limit $\mu \rightarrow 0$.

\begin{figure}[t]
\includegraphics[width=\columnwidth]{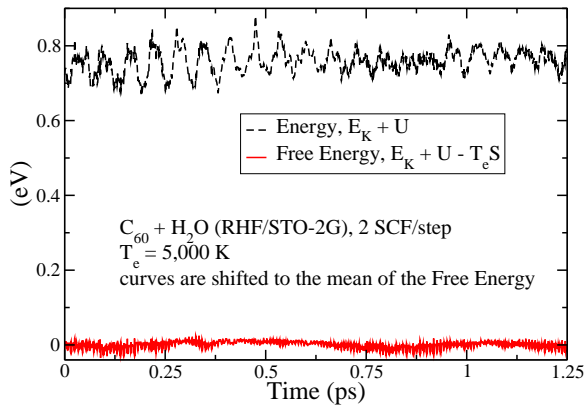}
\caption{\label{Fig_Bucky_water_T2500K}
\small (Color online) The fluctuations in the nuclear kinetic and potential energy, $E_{\rm
K} + U$, in comparison to the total free energy, $E^{\rm tot}_F = E_{\rm K} +
U - T_e{\cal S}$, for a density matrix extended Lagrangian free energy
molecular dynamics simulation of a single water molecule embedded within a
$C_{60}$ Bucky ball using Hartree-Fock theory.  The average nuclear
temperature was approximately $T_{\rm ion}\approx 1000$ K.}
\end{figure}

\begin{figure}[t]
\includegraphics[width=\columnwidth]{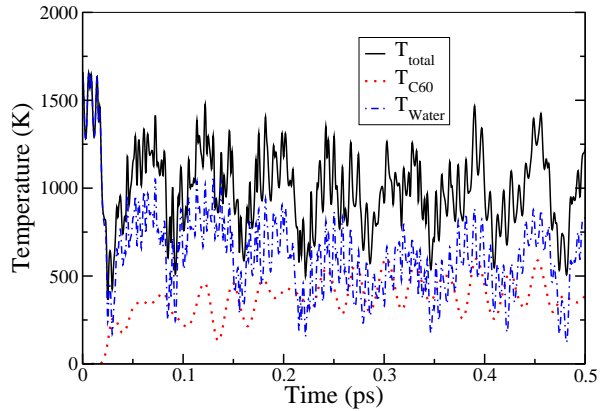}
\caption{\label{Fig_Bucky_water_T2500K_temperature} \small (Color online) The fluctuations in
the total nuclear temperature, $T_{\mathrm{total}}$, in comparison with the
temperature of only the Bucky ball, $T_{\mathrm{C60}}$ and the temperature of
the water molecule, $T_{\mathrm{water}}$. We start our simulation with the
Bucky ball at $T_{\mathrm{C60}} = 0$ K and a water molecule at the center
inside the Bucky ball with a non-zero initial velocity. While the total
temperature remains fairly constant, the energy transfer from the water
molecule to the Bucky ball is clearly visible in the rising
$T_{\mathrm{C60}}$ and the falling $T_{\mathrm{water}}$.}
\end{figure}

Figure~\ref{Fig_Bucky_water_T2500K} shows the corresponding behavior of the
free energy fluctuations of the sum of the nuclear kinetic and potential
energy, $E_{\rm K} + U$ for a single water molecule embedded in
a C${}_{60}$ Bucky ball. The electronic temperature was set to $T_e = 5,000$ K,
while the nuclear temperature of the Bucky ball was initially zero, and the
water molecule was given an initial velocity and hence had a non-zero initial
temperature. The time evolution of the total temperature, $T$, the temperature
of only the Bucky ball, $T_{\mathrm{C60}}$, and the temperature of only the
water molecule, $T_{\mathrm{water}}$, are shown in
Figure~\ref{Fig_Bucky_water_T2500K_temperature}. Clearly visible in the figure
is the process of kinetic energy transfer from the water molecule to the Bucky
ball; as the water molecule repeatedly collides with the wall of the Bucky
ball, kinetic energy is transferred between the two sub-systems and the Bucky
ball heats up, while the water molecule cools down. 

\begin{figure}[t]
\includegraphics[width=\columnwidth]{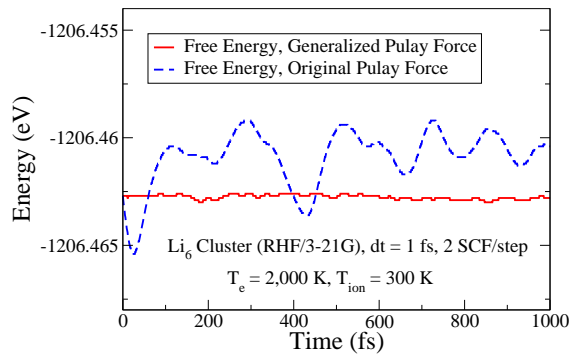}
\caption{\label{FE_GPulay}
\small (Color online) The fluctuations in the total free energy, $E^{\rm tot}_F = E_{\rm K} + U - T_e{\cal S}$, for a small Li cluster
at $T_e = 2,000$ K simulated using FreeON within the density matrix extension. 
The Pulay force has been calculated either with the original zero-temperature
expression, Eq.\ (\ref{PForce}), or its finite temperature generalization, Eq.\ (\ref{GPForce}).}
\end{figure}

If we use the original expression of the Pulay term \cite{PPulay69}, 
\begin{equation}\label{PForce}
{\cal F}^{\rm P} = 2\mathrm{Tr}[{\widetilde D}{\widetilde F}{\widetilde D}(\partial S/\partial R_k)],
\end{equation}
which is valid only for an idempotent density matrix 
in the limit of $T_e = 0$, i.e. when ${\widetilde D} = {\widetilde D}S{\widetilde D}$, we find that it affects
the energetics even at fairly modest electronic temperatures. Figure \ref{FE_GPulay}
gives a comparison of the free energy calculated either with the original zero-temperature
expression in Eq.\ (\ref{PForce}) or its finite temperature generalization in Eq.\ (\ref{GPForce}).
Only the generalized Pulay force has the correct variational properties derived
from the extended free energy Lagrangian, which gives a significant improvement 
in the conservation of the free energy.

\subsubsection{Plane wave extension}
 
\begin{figure}[t]
\includegraphics[width=\columnwidth]{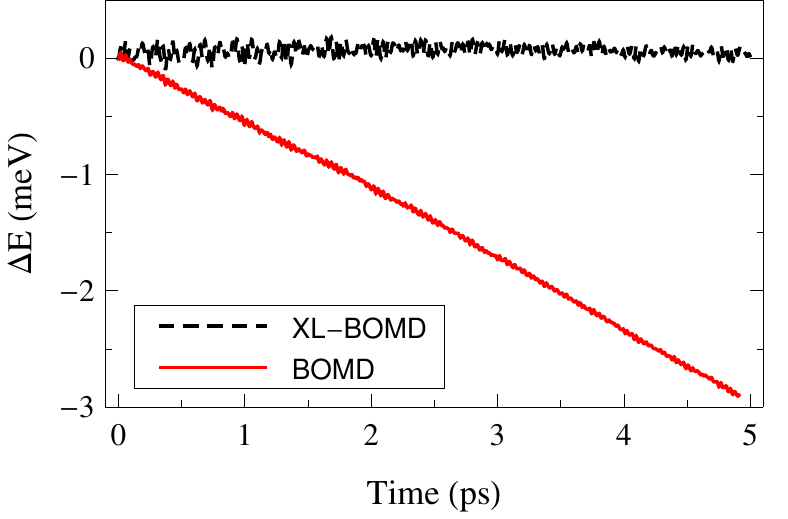}
\caption{\label{Fig_WF_Total_Energy}
\small (Color online) The change in total free energy as a function of time for a periodic 8 atom silicon
cell for conventional Born-Oppenheimer MD as implemented in VASP and extended Lagrangian 
free energy MD. Both simulations using a time step of 0.25 fs, and a energy convergence criteria of 0.5 meV.
The conventional Born-Oppenheimer MD show a clear unphysical systematic drift in comparison 
to the extended Lagrangian free energy MD.}
\end{figure}

\begin{figure}[t]
\includegraphics[width=\columnwidth]{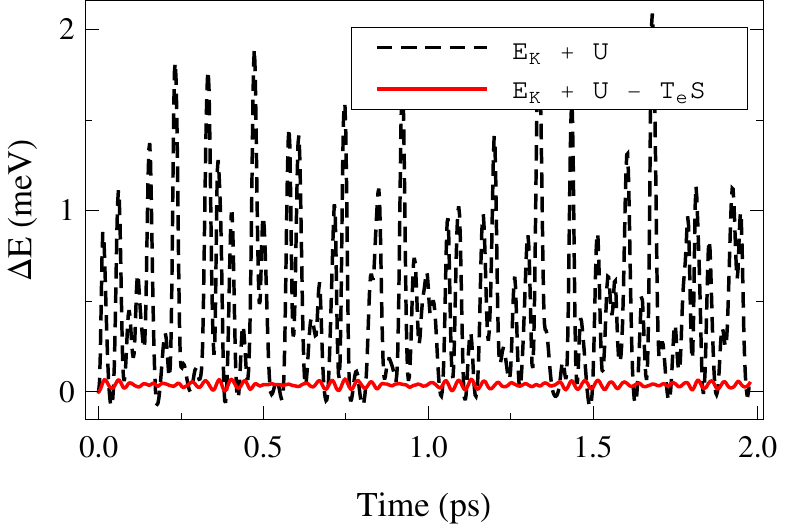}
\caption{\label{Fig_WF_ETotal_EFree}
\small (Color online) The fluctuations for the total nuclear and potential energy, $E_K + U$,
in comparison to the total free energy, $E_K +U +T_e S$, of a periodic 8 atom
silicon system. The simulation was carried out using extend Lagrangian free
energy MD in VASP. Using a time step of 0.25 fs and energy convergence criteria
of $5\times 10^{-10}$ eV with ultrasoft pseudopotentials. Both the nuclear and
electronic had an average temperature of about 500 K.}
\end{figure}

\begin{figure}[t]
\includegraphics[width=\columnwidth]{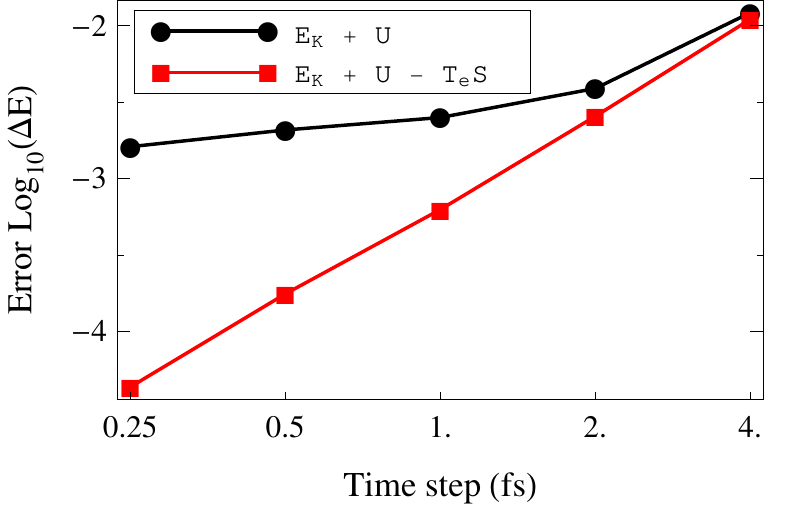}
\caption{\label{Fig_WF_Fluctuations}
\small (Color online) The error amplitudes as a function of time step for the total nuclear and potential energy, $E_K + U$, 
and the total free energy, $E_K +U +T_e S$, of a 8 atom periodic silicon cell}
\end{figure}

Figure~\ref{Fig_WF_Total_Energy} demonstrates how the total free energy for the conventional 
Born-Oppenheimer molecular dynamics and the plane wave extended Lagrangian free energy molecular dynamics
behave over time in the plane wave formulation.  A system of 8 silicon atoms are simulated for a total of
5 ps using a time step of 0.25 fs. The figure effectively depicts how the regular Born-Oppenheimer 
molecular dynamics exhibit an unphysical systematic drift in the total free energy. This issue is solved
by the use of the extended Lagrangian free energy formulation, which with equivalent settings otherwise
shows no systematic drift. 
In Fig.\ \ref{Fig_WF_ETotal_EFree} we show the same behavior for the plane wave formulation as in
Fig.\ \ref{Fig_water_T10000K} for the density formulation, where we demonstrate that the total free energy
is a constant of motion.
Figure \ref{Fig_WF_Fluctuations} shows the error as measured by the amplitude of the total energy vs. the free energy fluctuations
for various lengths of the integration time step $\delta t$. Only the free energy has the correct scaling with respect to
the time step.

\subsection{Fermi operator expansion}

\begin{figure}[t]
\includegraphics[width=\columnwidth]{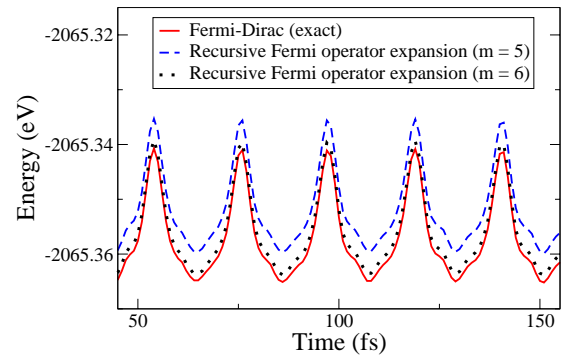}
\caption{\label{Fig_water_Recursion_m}
\small (Color online) The fluctuations in the total free energy, $E^{\rm tot}_F = E_{\rm K} + U - T_e{\cal S}$, for a water molecule, H$_2$O (PBE0/3-21G),
using either the exact Fermi-Dirac distribution or the recursive Fermi
operator expansion, Algorithm~\ref{Fermi_Op_Exp},
with 5 ($m=5$) or 6 ($m=6$) recursion steps. The simulation is based on a hybrid density functional (PBE0 \cite{CAdamo99}) for $T_e = 10,000$ K 
and $T_{\rm ion} \approx 300$ K using 2 SCF cycles per time step, $\delta t = 0.5$ fs, within the density matrix extension.}
\end{figure}

\begin{figure}[t]
\includegraphics[width=\columnwidth]{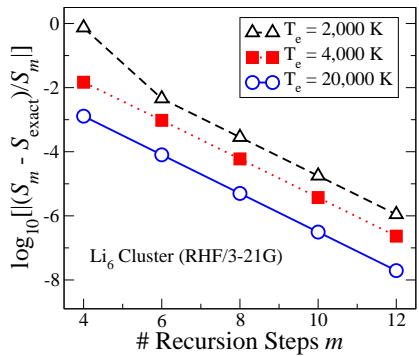}
\caption{\label{m_T_exp}
\small (Color online) The relative error in the entropy as a function of the number of expansion steps $m$ in the recursive 
Fermi operator expansion, Algorithm~\ref{Fermi_Op_Exp}, for three different electronic temperatures $T_e$.}
\end{figure}

The temperature-dependent density matrices for the simulations
in Fig.\ \ref{Fig_water_T10000K} were constructed using the ``exact" analytic Fermi-Dirac distribution of the states.
If we instead use the recursive Fermi operator expansion of Algorithm~\ref{Fermi_Op_Exp} we find no
significant deviation compared to the Fermi-Dirac result when we use more than 5 steps in the recursion.
Figure \ref{Fig_water_Recursion_m} shows the free energy calculated with the exact result vs. 
the recursive expansion. The example corresponds to the simulation of the lower panel in Fig.\ \ref{Fig_water_T10000K}. 
Using 5 recursion steps gives only a small constant shift
in the free energy. The accuracy for a specific expansion order $m$ will depend on the temperature $T_e$.
In Fig.\ \ref{m_T_exp} we show the error in the entropy as a function of $m$ for three different 
temperatures for a small Li cluster. In this case we find that the relative accuracy is reduced
for lower temperatures, though the rate of convergence as a function of the number of expansion steps $m$ is the same.

\section{Summary and Discussion}

Free energy generalizations of regular Born-Oppenheimer molecular dynamics for simulations 
at finite electronic temperatures are currently widely used \cite{MWeinert92,RWentzcovitch92,GKresse96,NMarzari97}.
In this paper first principles free energy molecular dynamics schemes
were developed based on extended Lagrangian Born-Oppenheimer molecular dynamics \cite{ANiklasson06,ANiklasson08,ANiklasson09,PSteneteg10}. 
The formulation was given both for density matrix and plane wave representations. The density matrix formulation
has numerically convenient expressions for the electronic forces as well as a recursive Fermi operator
expansion that are well suited for reduced complexity calculations.  Our extended Lagrangian free energy molecular dynamics 
enables efficient and accurate molecular dynamics simulations based on Hartree-Fock and density functional
theory (or any of their many extensions) both for localized orbital and plane wave pseudopotential calculations
with a temperature-dependent electronic degrees of freedom
and nuclear trajectories that evolve on the self-consistent free energy potential surface.

\section{Acknowledgment}

The Los Alamos National Laboratory is operated by Los Alamos National
Security, LLC for the NNSA of the USDoE under Contract No.
DE-AC52-06NA25396.  We gratefully acknowledge the support of the US Department of Energy through 
the LANL LDRD/ER program and the Swedish Foundation for Strategic Research (SSF) via 
Strategic Materials Research Center on Materials Science for Nanoscale Surface Engineering (MS2E),
 and The Göran Gustafsson Foundation for Research in Natural Sciences and Medicine for this work.  
Discussions with Marc Cawkwell, Erik Holmstr\"{o}m, Igor Abrikosov and Anders Odell 
as well as support from Travis Peery at the T-Division Ten-Bar Java Group are gratefully acknowledged.

\appendix

\section{Recursive Fermi operator expansion}
\label{appendix:recursive_operator_expansion}

There are at least 19 different (dubious) ways to calculate matrix exponentials \cite{CMoler03}. Consider
\begin{equation}
e^x = \left(e^{x/n}\right)^n =  \left( \frac{e^{x/(2n)}}{e^{-x/(2n)}}\right)^n,
\end{equation}
which after a first order Taylor expansion gives
\begin{equation}
e^x = \lim_{n \rightarrow \infty} \left( \frac{2n+x}{2n-x}\right)^n.
\end{equation}
This means that the Fermi-Dirac distribution function, ${\Phi}(x)$, can be approximated as
\begin{equation}
{\Phi}(x) = [e^x + 1]^{-1} = \lim_{n \rightarrow \infty} \frac{(2n-x)^n}{(2n+x)^n + (2n-x)^n},
\end{equation}
such that
\begin{equation}
{\Phi}(2n - 4nx) \approx \frac{x^n}{x^n + (1-x)^n}
\end{equation}
for large values of $n$.  The function
\begin{equation}
{f}_n(x) = \frac{x^n}{x^n + (1-x)^n}
\end{equation}
has the important recursive property,
\begin{equation}
{f}_{m \times n} (x) = {f}_m({f}_n(x)),
\end{equation}
which enables a rapid high-order expansions.
The following recursive approximation can therefore be used:
\begin{eqnarray}
{\Phi}\left[\beta(\varepsilon_i-\mu)\right] & = & {\Phi}(2n - 4nx_i) \nonumber \\
\approx {f}_{n}(x_i) & = & {f}_2({f}_2( \ldots {f}_2(x_i) \ldots )),
\end{eqnarray}
where
\begin{equation}
x_i = \frac{1}{2} - \frac{\beta}{4n}(\varepsilon_i -\mu)
\end{equation}
and with the recursion repeated $m$ times, i.\ e.\ for $n = 2^m$.

The density matrix at finite electronic temperatures,
\begin{equation}
D = \left[e^{\beta(H-\mu I)} + 1\right]^{-1} = {\Phi}\left[\beta(H-\mu I)\right],
\end{equation}
can thus be calculated using the recursive grand canonical Fermi operator expansion,
\begin{equation}
D = {f}_2\left(\ldots f_2\left({f}_2\left(\frac{1}{2}I - 2^{-(2+m)}\beta(H-\mu I)\right)\right) \ldots\right),
\end{equation}
which is generated by Algorithm~\ref{Fermi_Op_Exp}.

\section{Constant energy integration: a practical guide}

\subsection{Underlying approximations}

Even if a correct geometric and long-term energy conserving integration scheme, such as the Verlet algorithm, is used
both for the nuclear and the electronic degrees of freedom, a systematic energy drift may still appear in a simulation.
The most probable cause is the approximate formulation of the underlying dynamics. Any self-consistent first principles
calculation involves many different numerical approximations, of which some can be physically
motivated, others not. In the ideal case we would perform all calculations numerically exact, but for
an {\em approximate} underlying dynamics that is sufficiently close to our effective single particle theory.
The dominating major approximations due to incomplete SCF convergence and the finite time steps are 
treated within the extended Lagrangian framework and the geometric integration presented here. Other
approximations, such as truncated multipole expansions, finite arithmetics, thresholding 
of small matrix elements, and finite grid errors, have not been considered. First principles electronic structure codes have
in general never been able to generate exact energy conserving molecular dynamics. To obtain a well working
scheme that is stable and energy conserving under approximate SCF convergence may thus involve extensive prior 
testing, tuning and debugging, even if the dissipative electronic force discussed above is applied. In this way,
an energy conserving extended Lagrangian free energy (or Born-Oppenheimer) molecular dynamics simulation is an excellent test and signature
of a well working code.

\subsection{Thermostats}

In regular Born-Oppenheimer molecular dynamics simulations, various forms of thermostats are applied to generate, for example,
stable constant temperature (NVT) or constant pressure (NVP) ensembles. In this case the problem with a systematic energy drift
may not be seen directly, but the underlying dynamics is, of course, still wrong, even if a hypothetical ``exact'' 
thermostat could be used. By using an ever increasing number of SCF cycles the underlying energy drift can be kept low, 
but this is computationally expensive. Thermostats typically act through a rescaling of the forces or the velocities of the nuclear degrees of freedom.
Unfortunately, this approach may lead to an unphysical energy transfer between high-frequency and low-frequency modes \cite{SCHarvey98}.
The problem is particularly significant when there is a large difference between the atomic masses \cite{DFukushi00}.

\subsection{Practical guide}

Depending on the accuracy of the electronic structure code used in the molecular dynamics simulation, we give the
following practical guide for a constant free energy simulation: 1) Use the extended Lagrangian formulation \cite{ANiklasson08} 
and a geometric integration scheme 
\cite{LVerlet67,BJLeimkuhler04,ANiklasson08,AOdell09} for the integration of both the nuclear and the electronic degrees of freedom. 
2) If the calculation is noisy, use electronic dissipation
\cite{ANiklasson09}, e.g. as in Eq.\ (\ref{VRL_Damp}). 3) If the simulation
still shows excessive fluctuations in the energy or a systematic drift we
suggest a careful review and tuning of the numerical approximations in the
underlying electronic description. 4) If the energetics still is unstable, and
then only as a last resort, one could use a stabilizing velocity or force
rescaling. In that case we suggest the following velocity rescaling, which is
constructed to minimize the perturbation of the molecular trajectories:
\begin{equation}
c_n = \sqrt{1 + \frac{\delta t}{\tau}\left(\frac{E_F^0 - E_F^{\rm tot}(n)}{E_{\rm K}(n)}  
+ \gamma \sum_{j=1}^{n-1} \frac{E_F^0 - E_F^{\rm tot}(j)}{E_{\rm K}(j)}\right)},
\end{equation}
\begin{equation}
 {\bf v}_i \rightarrow c_n {\bf v}_i.
\end{equation}
Here ${\bf v}_i$ is the velocity of particle $i$, which is rescaled by the constant $c_n$ at time step $n$. $E_F^0$ is the
desired target (free) energy, $E_F^{\rm tot}(n) = E_{\rm K} + U - T_e{\cal S}$ is the total (free) energy and
$E_{\rm K}(n)$ is the nuclear kinetic energy at time step $n$. 
The constant $\tau$ is a chosen relaxation time and $\gamma$ is a weight factor for
the integrated error term. The rescaling above is loosely based on the Berendsen thermostat \cite{HJCBerendsen84} including an 
integrated error term (if we chose $|\gamma| > 0$) as
is often used in conventional control theory. The velocity rescaling factor $c_n$ is typically small and if the Verlet integration scheme is used, $c_n$
scales with the length of the integration step $\delta t$ as $c_n = 1 + {\cal O}(\delta t^3)$ or $c_n = 1 + {\cal O}(\delta t^2)$, 
depending on how the relaxation time $\tau$ is chosen.  For higher-order integration schemes, 
$c_n = 1+ {\cal O}(\delta t^k)$ for larger integers of $k$.  This means that even if the velocity rescaling is unphysical, the
error can at least be kept very small, with only minor perturbations of the molecular trajectories. 
The velocity rescaling above was not necessary for any of the simulations performed for this article.

\newpage

\newpage

\newpage

\newpage

~                             

\end{document}